\documentclass[aps,prx,twocolumn,superscriptaddress,showpacs]{revtex4-2} 
\usepackage{amssymb}
\usepackage{suffix}
\usepackage{mathtools}
\usepackage[utf8]{inputenc}
\usepackage{booktabs}
\usepackage{cases}
\usepackage[multiple]{footmisc}
\usepackage{dcolumn}
\usepackage{color,soul}
\usepackage{rotating}
\usepackage{perpage}
\usepackage{xcolor}
\usepackage{soul}
\usepackage{tikz}
\usepackage[T1]{fontenc}
\usepackage{etoolbox}
\usepackage{graphics}
\usepackage{siunitx}
\usepackage{hyperref}
\usepackage{float}
\usepackage{multirow}
\usepackage{mathtools}
\usepackage{bm}
\usepackage{url}

\usepackage{tikz}
\usepackage{tikz-3dplot}
\usepackage [outdir=./]{epstopdf}
\usepackage{setspace}

\newcommand{\SD}[1]{\textcolor{blue}{\fbox{SD} {\sl#1}}}

\begin{document}
\title{Engineering of topological phases in driven thin topological insulator:\\ Structure inversion asymmetry effect}

\author{S. Sajad Dabiri}
\address{Department of Physics, Shahid Beheshti University, G.C., Evin, 19839-63113 Tehran, Iran}
\author{Hosein Cheraghchi}
\email{cheraghchi@du.ac.ir}
\address{School of Physics, Damghan University, P.O. Box 36716-41167, Damghan, Iran}
\address{School of Physics, Institute for Research in Fundamental Sciences (IPM), 19395-5531, Tehran}

\date{\today}
\vspace{1cm}
\newbox\absbox
\begin{abstract}
We investigate the effect of a high frequency electromagnetic field with both of circularly and linearly polarization, on the emergence of quantum phases on thin topological insulators. Simultaneously, the influence of the system parameters (such as magnetic impurity, thickness engineering and structural inversion asymmetry of the potential) on emergence of topological phases is studied. We take our attention to the high frequency regime in which it is possible to consider an expansion for the Floquet Hamiltonian in terms of orders of $1/\Omega$. The topological invariants are determined and it is demonstrated that some phase transitions between quantum anomalous Hall insulator(QAHI), quantum pseudospin Hall insulator (QPHI), quantum spin Hall insulator (QSHI) and normal insulator (NI) can be induced by altering the aforementioned parameters of the system. To avoid heating process, tuning of the system parameters gives us the opportunity to observe these phase transitions at small intensities of light.
\end{abstract}
\maketitle

\section{Introduction}
Since the discovery of Quantum Hall effect \cite{klitzing} and its theoretical description based on topological argument \cite{thouless}, there have been a lot of interests and investigations about topological phases\cite{hasan,sczhang,ando}. In particular, some materials with strong spin-orbit coupling such as $Bi_2Se_3$ and (Bi,Sb)$_2$ Te$_3$ family materials were identified to be 3D topological insulators. These kinds of materials support gapless surface states with dispersion of Dirac cone on their surfaces. In the thin film version of these materials, two surface wave functions overlap with each other giving rise to an effective two dimensional insulator. Thin film geometry has the benefit of suppressing the bulk conduction in favor of observing the quantized conduction of edge states, moreover in this geometry, the chemical potential is more controllable. On the other hand by changing the thickness, one can control the topology of the system \cite{thick1,thick2}. The quantum anomalous Hall effect (QAH) which is a quantum Hall effect without external magnetic field was observed in TI thin films which are doped by magnetic impurities, such as Cr-doped \cite{science340,prl113137201,nature10731} and V- doped \cite{qahv} $(Bi,Sb)_2Te_3$ . There is a complete classification of topological phases of noninteracting fermionic systems according to their symmetries known as Altland-Zirnbauer (AZ) table \cite{zirnbauer,altland,ryu}. 

Since the past decade, because of developments in technology of probing and controlling out-of-equilibrium systems, a lot of attention have been attracted to time-periodic systems and Floquet engineering; i. e. inducing different behaviours in materials by driving them periodically in time \cite{demand}. Topological effects have been shown to appear in primary insulating \cite{tanaka,2011Lindner,2013Gomez,katsnelson,ezawasilicene,shelykh} or semimetallic materials \cite{oka,kitagawa2011,driven,fweyl} by application of periodic driving fields. In particular, switching between quantum spin Hall insulating phase (QSHI) and normal insulating phase (NI) has been shown to emerge in HgTe quantum wells, just by means of an illumination of off-resonant linear polarized light \cite{shelykh}. 

Indeed, Floquet theorem plays an essential role in studying time periodic systems \cite{shirley,sambe}. Intriguingly the Floquet-Bloch states were observed on the surface of a three dimensional topological insulator illuminated by circular and linear polarized field\cite{obs,mahmood}. Recently, in the absence of applied magnetic field, the anomalous Hall effect in graphene is observed while irradiated by circularly polarized mid-infrared laser pulse\cite{mciver}. The realization of periodic driven systems can also be observed in cold atoms by periodically changing the laser field of optical lattices \cite{cold1,cold2,cold3}.

The complete classification of non-interacting time-periodic topological phases and their corresponding invariants has been introduced in Refs. \onlinecite{roy,yao2017}. In fact, the usage of high frequency drive which is larger than any energy scale of the system has some benefits: the first, the topological properties of such driven systems resembles the static ones \cite{2013Gomez,rudnerrev}; and the second, heating process is more suppressed for such high frequency off-resonant drive and finally the third, the occupations of states are similar to the equilibrium ones \cite{mitra4}. As a result, the conductance of edge states in a topological phase will be approximately quantized. In the off-resonant regime, the system does not host "anomalous edge states" (the edge states which appear in the dynamical gaps of quasienergy spectrum) as stated in Ref.\cite{rudner2013}, so it is possible to consider only the zero-energy gap in the first Floquet zone.

In this paper, we investigate the influence of high frequency circularly and linearly-polarized light on electronic properties of a topological insulator thin films, along with the influence of magnetic impurities, thickness engineering and structure inversion asymmetry (SIA) potential. In our past work \cite{dabiri}, we considered the effect of circularly polarized light on topological phases of topological insulator thin film in the aforementioned parameters except SIA potential which is in our focus in this paper. As a consequence of substrate \cite{nature584}, emergence of SIA potential in experiment is inevitable. In the absence of the drive, it has been shown that the SIA potential is able to change the topology of the system \cite{electrically,xing2012}. Also this term breaks the pseudospin symmetry which has been assumed to be conserved in our previous work \cite{dabiri}. At the end, we try to derive The Floquet Hamiltonian driven by linearly-polarized light at high frequency regime. This polarization does not break time-reversal symmetry however, such a pseudospin symmetry is absent. It is shown that by the application of structure inversion asymmetry, one observes a phase transition from QSHI to NI phase. 

The paper is organized as follows: In Sec. \ref{S2}, we derive the low energy Hamiltonian of two-dimension topological insulator by means of high-frequency expansion formalism \cite{Bukov,eckart,bw}. In Sec. \ref{S3}A , firstly we present the Floquet Hamiltonian when SIA potential is zero and then in Sec. \ref{S3}B, we will turn on the SIA term and extract its Floquet Hamiltonian. The phase diagram and topological invariants of this effective Hamiltonian is also investigated in this section. There is a discussion about the calculation of pseudo-spin Chern number at the end of this section. The effect of linearly polarized light on the system along with $V_{sia}$ is also discussed in Sec.\ref{S4}. We present summary and conclusion in Sec.\ref{S5}. In Appendix. \ref{A1}, the derivation of the Floquet Hamiltonian driven by a linearly polarized light is presented. At the end, there is also a comparison between the results originated from the perturbative expanded Hamiltonian and the exact Floquet Hamiltonian in Appendix. \ref{A2}.

\section{Model and formalism}\label{S2}
\emph{Dark Hamiltonian:}
The low energy effective Hamiltonian for $Bi_2Se_3$ and (Bi,Sb)$_2$ Te$_3$ family materials \cite{nature2009}, which are well-known 3D topological insulators, is dominated by the surface states. These surface states are composed of gappless Dirac cones at $\Gamma$ point. In thin-film's geometry, the two Dirac cones can be hybridized giving rise to open a gap and make the system an effective 2D insulator. On the other hand, as a result of doping with magnetic impurities such as Ti, V, Cr and Fe, the quantum anomalous Hall phase may appear. However as we will see in the following, even though there is no any magnetic doping, the light illumination can still lead to emergence of QAHI phase. Neglecting the particle-hole asymmetry term which does not change the topological behaviour of the system, the low energy effective Hamiltonian of the surface states around $\Gamma$ point is written as the following \cite{effective,yu}.

\begin{equation}
\begin{aligned}
\centering
&H^{\text{dark}}(\textbf{k})=
\hbar {v}_f \tau_z\otimes (k_y \sigma _x -k_x \sigma _y) 
+\Delta (\textbf{k})\tau_x\otimes \sigma _0 
\\&+V_{sia}\tau_z\otimes \sigma _0 
+ M_z \tau_0\otimes\sigma_z 
\end{aligned}
\label{eq:firsthamil}
\end{equation}
In the basis set of $|t,\uparrow \rangle ,|t,\downarrow \rangle ,|b,\uparrow \rangle ,|b,\downarrow \rangle$, where $t (d)$ denotes to the top (bottom) surface states and $\uparrow (\downarrow)$ refers to the up (down) spin states. $\tau_i (\sigma_i)$ are Pauli matrices in the surface (spin) space. 
The first term describes the surface states with Fermi velocity of $v_f$.
The second term comes from the tunneling between surfaces in which $\Delta(\textbf{k})=\Delta_0+\Delta_1 k^2$ for thin film of Bi$_2$ Se$_3$ and [(Bi,Sb)$_2$ Te$_3$] family but thinner than $ d=5 nm$\cite{nature584,prb81041307}. The parameters $\Delta_0$ and $\Delta_1 $ depend on the thickness of TI thin film which are fitted experimentally. The structural inversion asymmetry $V_{sia}$ in the third term, is due to perpendicular potential difference between two surfaces which may be induced by an applied electric field or substrate\cite{nature584}. The last term refers to perpendicular ferromagnetic exchange field which is originated from the magnetic impurities doped in Bi$_2$Se$_3$ and [(Bi,Sb)$_2$ Te$_3$] family\cite{Magnetic}. In the whole paper, those parameters which are taken in our numerical calculations would be as $\Delta_0=35$~meV and $\Delta_1=\pm10$~eV\AA$^2$ and the Fermi velocity is set to $v_f=4.48 \times 10^{5}$~m/s and the photon energy is considered to be $\hbar \Omega=1$~eV.

It is convenient to do a unitary transformation on the previous basis. By defining the bonding and antibonding states as $\vert \psi_b,\uparrow\downarrow\rangle=\left( \vert t,\uparrow\downarrow\rangle+\vert b,\uparrow\downarrow\rangle \right)/ \sqrt{2}$ and
$\vert \psi_{ab},\uparrow\downarrow\rangle=\left( \vert t,\uparrow\downarrow\rangle-\vert b,\uparrow\downarrow\rangle \right)/\sqrt{2}$, we write the Hamiltonian \ref{eq:firsthamil} in the new basis of $|\psi_b,\uparrow \rangle ,|\psi_{ab},\downarrow \rangle ,|\psi_b,\downarrow \rangle ,|\psi_{ab},\uparrow \rangle$. 
\begin{equation}
\begin{aligned}
H^{\text{dark}} (\textbf{k})=&\text{$\hbar $v}_f \left(k_y \tilde{\tau}_0\tilde{\sigma} _x- k_x\tilde{\tau}_z \tilde{\sigma} _y \right) \\
&+\Big(\Delta(\textbf{k}) \tilde{\tau}_0+ M_z \tilde{\tau}_z \Big)\tilde{\sigma} _z+V_{sia}\tilde{\tau}_x \tilde{\sigma}_x
\label{eq:hamiltotalchiral} 
\end{aligned}
\end{equation}

in which $\tilde{\tau}_i$ and $\tilde{\sigma}_i$, $(i=0,x,y,z)$ are Pauli matrices in a mixed space and also spin such that $\tilde{\tau}\otimes \tilde{\sigma}$ spans the new basis mentioned above.

If $V_{sia}=0$, the above effective Hamiltonian (Eq.\ref{eq:hamiltotalchiral}) commutes with $\tilde{\tau}_z$ and can be splitted into two blocked diagonal $2\times2$ matrices separated by a pseudo-spin index as the following
\begin{equation}
\centering
h^{\text{dark}}_{\alpha }(\textbf{k})=\text{$\hbar $v}_f \left(k_y \tilde{\sigma} _x- \alpha k_x \tilde{\sigma} _y\right)+\Big(\Delta(\textbf{k})+\alpha M_z \Big)\tilde{\sigma }_z
\label{eq:hamilchiral} 
\end{equation}
where $\alpha=\pm$ is the pseudo-spin index. Therefore, in this case, Hamiltonian would be pseudo-spin polarized which its trace can appear in some selective rules in quantum transport through magnetically doped topological insulator's film\cite{sabze}.

\emph{Photon-dressed Hamiltonian:}
A circularly-polarized light with the wave vector perpendicular to the thin film surface can be described by the vector potential $\overrightarrow{A}(t)=A_0(\sin(\Omega t),\cos(\Omega t))$. Where $|\Omega|=2 \pi /T$ is the frequency of light and $T$ is the period of vector potential, $A(t+T)=A(t)$. The sign of $\Omega$ determines the right or left-handed polarization. After irradiation of light, the wave-vector in Hamiltonian \ref{eq:firsthamil}, is changed to $k_i \longrightarrow k_i+\frac{e A_i}{\hbar}$. 

Generally one should use the Floquet approach for time-periodic systems. However, in case, energy of photons assisted in Floquet Hamiltonian is much higher than any characteristic energy of the system (band width), the system evolves according to an effective and static Hamiltonian which describes the evolution of Floquet state at the stroboscopic times i.e. the integer multiple of period $T$, which is definable through $H_{eff}=\frac{i}{T}log(\mathcal{T}exp(-i\int_0^TH(t)dt))$, where $\mathcal{T}$ denotes the time ordering operator. This effective Hamiltonian has a series expansion in terms of the inverse frequency \cite{Bukov,eckart,bw}

\begin{equation}
H=H^0+\sum_{m}(m\hbar \Omega)^{-1}[H^{-m},H^{+m}]+O(1/(\hbar\Omega)^2)
\label{photon_H}
\end{equation}
where in commutation relation, operators are defined as $H^{ m}=1/T \int_{0}^{T} H(t) e^{ im |\Omega| t} dt$. One can check that for our Hamiltonian $H^{i}=0$ (for $i\ne 0,\pm1$). 

The high frequency behavior of periodic systems is well described generally in Refs. \cite{Bukov,eckart,bw}. In high frequency regime, to derive effective and static Hamiltonian, there are some developed expansions such as vanVleck (vV) \cite{Bukov,eckart}, Floquet-Magnus (FM) \cite{magnus} and Brillouin-Wigner (BW) expansions \cite{bw}. In Eq.\ref{photon_H}, we have used the van-Vleck or BW expansion which yields the same result up to the first order   and does not depend on the phase of the drive as opposed to the FM expansion. The dependence of FM expansion on the phase of the drive leads to spurious symmetry breaking of energy bands which is not desirable \cite{eckart}.

The laser intensity is described by dimensionless parameter $\mathcal{A} a_0$ where $\mathcal{A}=eA_0/\hbar$ and $a_0\approx 4 $ ~\AA is typical lattice constant (being of the order of angstrom). We assume that this quantity is much smaller than unity, so we have $\mathcal{A}\ll 1/a_0$. Then for our system near the $\Gamma$ point \cite{kitagawa2011,ezawasilicene}, we can write 
\begin{equation}
H=H^0+(\hbar \Omega)^{-1}[H^{-1},H^{+1}]+\mathcal{O}(\frac{1}{(\hbar\Omega)^2})
\label{one_photon_H}
\end{equation}
Thus one-photon processes are dominant. Let us first review what happens if SIA potential is zero, although in experiment, because of substrate-induced potentials, the presence of SIA potential is inevitable. 

\section{Results} \label{S3}
\subsection{Circular Polarization}
\subsubsection{Perpendicular Magnetization}\label{S3_a}
To derive the Floquet Hamiltonian in the presence of high-frequency light illumination, we should calculate the Fourier component of the Hamiltonian. In the absence of $V_{sia}$, the dark Hamiltonian decouples into two $2\times2$ matrices. After the Peierls substitution of the wave vector in Eq.\ref{eq:hamilchiral} and also taking the integrals and substitution in Eq. \ref{one_photon_H}, we obtain 
\begin{equation}
h^{Driven}_\alpha=\hbar \eta_\alpha v_f ( k_y \tilde{\sigma}_x-\alpha k_x \tilde{\sigma}_y)+[\Delta'({\bf k})+\alpha(M_z+m_{\Omega})]\tilde{\sigma}_z
\label{Photo_Hamiltonian}
\end{equation}
in which
\begin{equation}
\begin{aligned}
&\Delta'(\textbf{k})=\Delta_0'+k^2\Delta_1 ~~ \Delta_0'=\Delta_0+\mathcal{A}^2\Delta_1, ~~ A^\prime=\frac{ \mathcal{A}^2 }{\hbar \Omega} \\
&\eta_\alpha=1- 2\alpha A^\prime \Delta_1 , ~~ m_{\Omega}=\hbar^2v_f^2A^\prime
\end{aligned}
\end{equation}
The phase diagram and also specific features of this Hamiltonian has been investigated in our previous work \cite{dabiri}. It was shown that pseudo-spin $\tilde{\tau}_z=i\sigma_z\tau_x$ (or equivalently mirror symmetry with respect to 2D plane) conservation leads to defining pseudo-spin Chern number as the following
\begin{equation}
\mathcal{C}_\alpha=\frac \alpha 2 \big (\text{sgn}\Delta_1 -\text{sgn} (\Delta_0'+\alpha m) \big)~~ \text{for}\,\ V_{sia}=0
\label{pseudo-spinchernnumber}
.\end{equation}
In the above formula, $m=M_z+m_{\Omega}$ where $m_\Omega$ is a mass term attributed to light illumination. The total Chern number was defined $\mathcal{C}=\mathcal{C}_++\mathcal{C}_-$ and total pseudo-spin Chern number $\mathcal{C}^\prime=1/2(\mathcal{C}_+-\mathcal{C}_-)$.
In the next section, we will turn on the structure inversion asymmetry term which breaks the pseudo-spin $\tilde{\tau}_z$ conservation and gives rise to a richer phase diagram.

\subsection{Structure Inversion Asymmetry} \label{S3_b}
The photo-induced Hamiltonian would be deduced as the following
\begin{equation}
\begin{aligned}
H=& (1-2A'\Delta_1 \tilde{\tau}_z) \hbar v_f ( k_y \tilde{\sigma}_x- k_x\tilde{\tau}_z \tilde{\sigma}_y) \\
& +[\Delta'({\bf k})+\tilde{\tau}_z m]\tilde{\sigma}_z +V_{sia}\tilde{\tau}_x\tilde{\sigma}_x
\label{Photo_Hamiltonian2}
\end{aligned}
\end{equation} 
As mentioned before, it would be important to notice that turning the SIA potential on, causes Hamiltonian to not commute with the pseudo-spin operator $\tilde{\tau}_z$. As a result, the quantum number $\alpha$ does not preserve when $V_{sia} \neq 0$. However, the pseudo-spin Chern number is still definable while there is no crossing between the eigenvalues of the pseudo-spin operator. \iffalse On the other hand, nontrivial pseudo-spin Chern number does not guarantee the existence of edge states. This is a negative expression. Because if we have no edge state why we choose this kind of Chern number for determining topological phases?}\SD{Our problem without light illumination is solved in paper (PHYSICAL REVIEW B 85, 045118 (2012)) where the pseudospin Chern number is used to describe phases. I calculated this quantity to see if there is a relation between edge states and nontrivial pseudospin Chern number.}\fi The dispersion relation for Hamiltonian \ref{Photo_Hamiltonian2} reads as
\begin{equation}
\begin{aligned}
E(k)&= \pm \Bigg((\hbar v_f k)^2(1+4A'^2{\Delta_1}^2)+(\Delta'(k))^2+m^2+V_{sia}^2\\
&\pm 2 \Big[ (\hbar v_f k)^2( V_{sia}^2+4A'^2\hbar^2k^2v_f^2{\Delta_1}^2)\\
&+ m^2\left(\left(\Delta'(k)\right){}^2+V_{sia}^2\right)-4A'\hbar^2k^2mv_f^2\Delta_1\Delta'(k) \Big] ^{\frac{1}{2}}\Bigg)^{\frac{1}{2}}
\label{eq:totalE} 
\end{aligned} 
\end{equation}
\begin{figure*}
\includegraphics[width=\linewidth,trim={0 0 0.25cm 0}, clip]{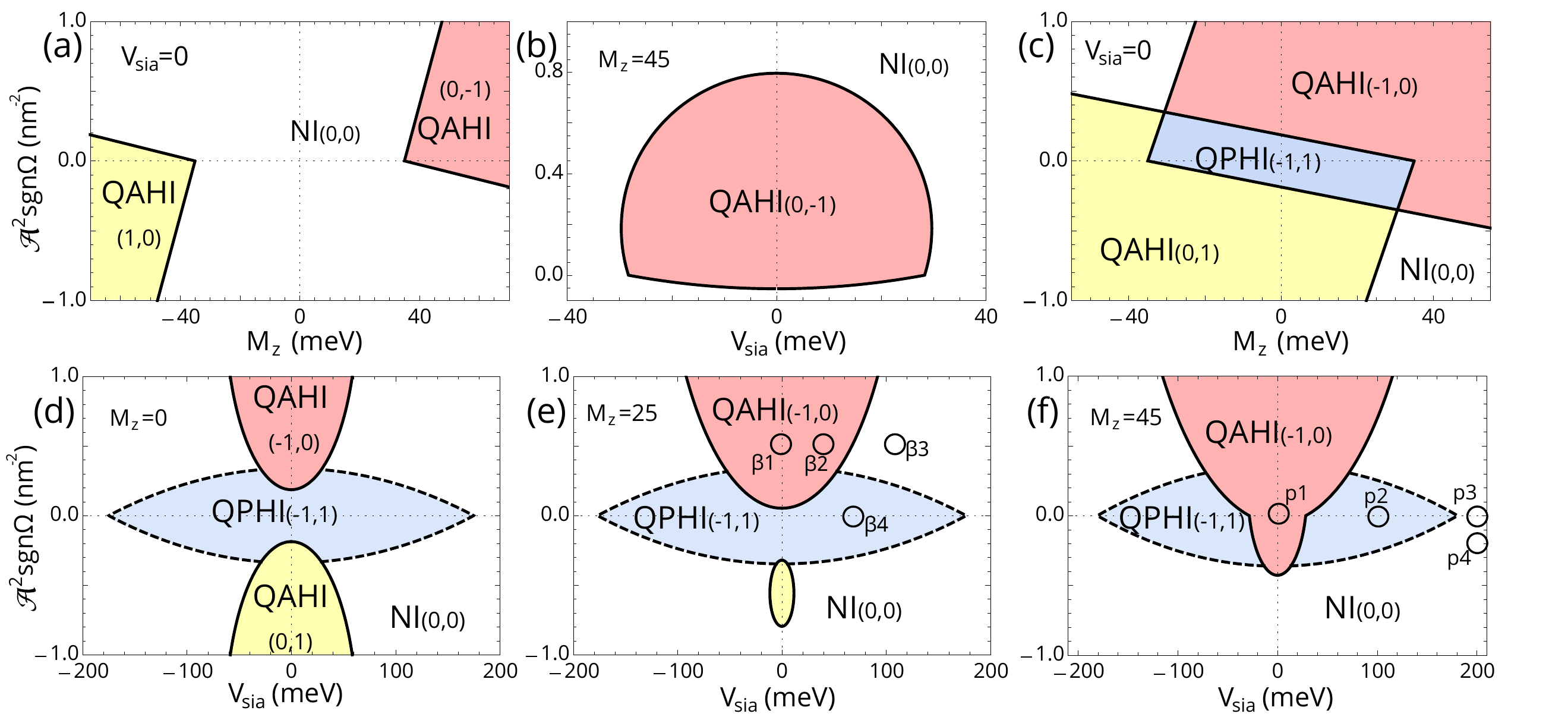} 
\caption{Evolution of the phase diagram with structural inversion asymmetric potential $V_{sia}$ for different values of parameters. (a-b) the case (i); with $\Delta_0\Delta_1>0$, (c-f) the case (ii); with $\Delta_0\Delta_1<0$. Solid (dashed) curves are related to the gap closing at (out of) the $\Gamma$ point. Blue (white) regions are phases with (without) counterpropagating edge states. Pseudospin Chern numbers $(\mathcal{C}_+,\mathcal{C}_-)$ are shown on each phase. }
\label{avdiagram}
\end{figure*}

At first, we look for the gap closing where the topology of the system can change. As a result of the above spectrum, the gap closing and so the phase boundaries occur at two conditions; the first one emerges at the $\Gamma$ point ($k=0$) if and only if the total mass term is equal to :
\begin{equation}
m^2=V_{sia}^2+{\Delta'}_0^2
\label{condition1_SIA}
\end{equation}
and also the second one occurs at $k^2=-\frac{\Delta'_0}{\Delta_1}-2A'm>0$ if and only if the following condition is satisfied:
\begin{equation}
V_{sia}^2=\frac{(-1+4A'^2\Delta_1^2)}{\Delta_1}(-m^2\Delta_1+\hbar^2v_f^2(\Delta'_0+2A'm\Delta_1))
\label{condition2_SIA}
\end{equation}

The phase diagram for different values of $M_z$ is displayed in Fig.\ref{avdiagram}. Figs. \ref{avdiagram} (a-b) correspond to the case (i) $\Delta_0\Delta_1>0$ and Figs. \ref{avdiagram} (c-f) correspond to the case (ii) $(\Delta_0\Delta_1<0)$. \emph{The gap closing at (or out of) the $\Gamma$ point is indicated by the solid (or dashed) lines}. 

It is notable that the gap closing at $\Gamma$ may lead to a band inversion and changing the Chern number. By integrating the Berry curvature of occupied bands or analysing the band inversion points, one obtains the total Chern number as follows
\begin{equation}
\mathcal{C}=-Sgn(m)\Theta(m^2-{\Delta'}_0^2-V_{sia}^2)
\label{chernnumber}
\end{equation}
In the above expression $Sgn(x)$ is the sign function and $\Theta(x)$ is the Heaviside step function.
The general feature of the case (i) as shown in Fig.\ref{avdiagram}(a) is composed of two QAHI regions which are separated by a NI phase, Fig.\ref{avdiagram} a(b) shows the phase diagram when $V_{sia}=0 (M_z=45)$. The pseudo-spin Chern number of QAHI phase occurring at $M_z>0$ is $(\mathcal{C}_+,\mathcal{C}_-)=(0,-1)$, while it is $(1,0)$ for QAHI in $M_z<0$ region. We describe the method of calculating pseudo-spin Chern number in the next section.

We call the phase with counter-propagating edge states, QPHI phase in which the edge states are fully pseudo-spin polarized provided that $V_{sia}=0$, however, for non-zero $V_{sia}$, these edge states are partially pseudo-spin polarized depending on the value of structure inversion asymmetry. The QPHI (NI) phase corresponds to the blue (white) region in Fig.\ref{avdiagram}, respectively. 

Phase diagrams for the case (ii) are indicated in Fig.\ref{avdiagram}(c-f); panel (c) is plotted if $V_{sia}=0$ while panels (d-f) are attributed to different values of $M_z$. In the case (ii), two QAHI phases appear, as well as QPHI and NI phases (denoted by the blue and white regions, respectively). In this case, the pseudo-spin Chern number is equal to $(\mathcal{C}_+,\mathcal{C}_-)=(-1,1)$ in QPHI and $(\mathcal{C}_+,\mathcal{C}_-)=(0,0)$ in NI phases (Fig.\ref{avdiagram}(c-f)). 

Let us conclude that in the case (i), by switching on the $V_{\text{sia}}$ potential, the system can tolerate a phase transition from QAHI into NI phase if SIA potential exceeds a critical value of $V^{Cr.}_{\text{sia}}$, Fig.\ref{avdiagram}(a-b). This critical value of potential depends on the light intensity and also polarization of light. 

In the case (ii), depending on the value of magnetization $M_z$, $V_{\text{sia}}$ potential play a crucial role to turn QAHI phase into QPHI and/or NI phases if it exceeds a critical value Fig.\ref{avdiagram}(c-f). In the non-magnetic case, depending on the polarization and intensity of light, SIA potential can control the phase transition from QAHI/QPHI to NI, however, presence of magnetization breaks symmetry of mass term in terms of light polarization. As drawn in Fig. \ref{avdiagram} f, under illumination of the right-handed polarization, by increasing SIA potential, one can observe a phase transition between QAHI and NI phases, while this phase transition is absent under illumination of the left-handed polarized light. As a generic property, depending on the value of SIA potential or magnetization, the phase diagram is asymmetric in terms of light polarization. It means that engineering of the transitions from QAHI to NI or QPHI phases, only happens by switching on the polarization of light. In small intensity of light illuminated on TI thin film with magnetic dopants, SIA potential can induce much richer phase transitions such as QAHI-QPHI-NI. This result is worthy because variation in SIA potential is much more accessible than the phase transitions induced by magnetization. Usually magnetization is set when once the sample is prepared, however, light parameters and also SIA potential are those parameters which are simply adjustable after sample fabrication. As we will explain in the experimental realization section, occurence of the phase transitions at small intensities, is favorite to decrease heating problems. 

Actually our previous results reported in Ref.\cite{dabiri} can be also rebuilt by setting $V_{\text{sia}}=0$ when we study the phase diagram in terms of $M_z$ and light parameters. The phase diagram in Fig. \ref{avdiagram}(f), in the absence of light illumination, is similar to the case reported in Ref. \cite{Magnetic} where switching the $V_{\text{sia}}$ on, leads to the phase transition from QAHI into QPHI and then to NI. Also our results are in good agreement with the phase diagrams presented in Refs.\cite{electrically,xing2012}

\begin{figure}
\includegraphics[width=0.85 \linewidth]{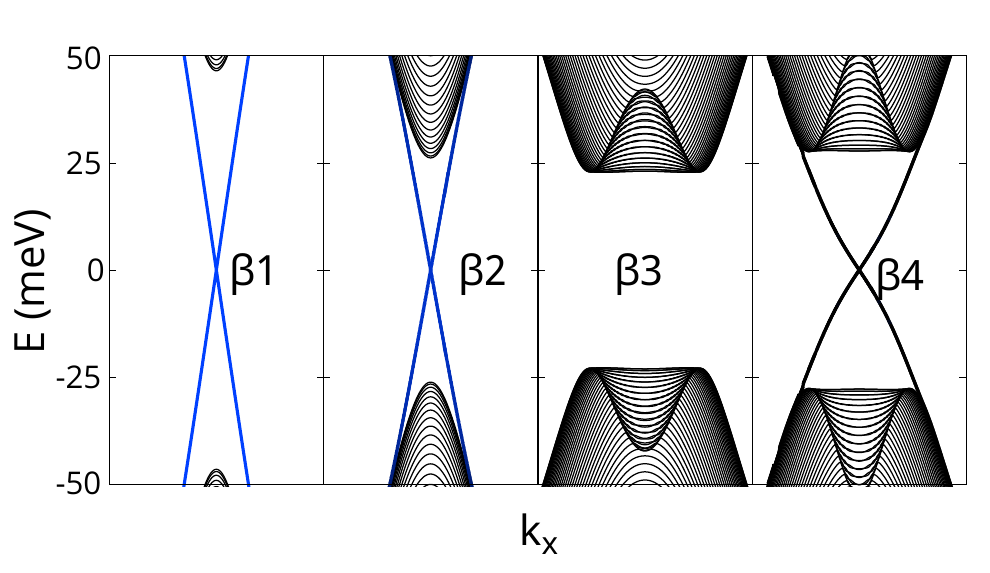} \includegraphics[width=0.1 \linewidth]{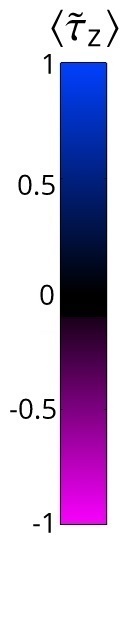} 
\caption{The energy bands of a nanoribbon TI thin film for the special phases depicted in Fig.\ref{avdiagram}(e) as $\beta_1$ to $\beta_4$. The nanoribbon width is 240nm and its unit cell length $a=2nm$. The color on edge states shows the expectation value of pseudospin operator $\langle\tilde{\tau}_z\rangle$.}
\label{betas}
\end{figure}

To make the story more clear, we take four points in Fig.\ref{avdiagram} (e) denoted by $\beta_1$ to $\beta_4$ and draw the dispersion relation of its nanoribbon version in Fig. \ref{betas}. The energy bands confirm each phase which is claimed in the phase diagram. The expectation value of the pseudo-spin operator $\langle\tilde{\tau}_z\rangle$ for the edge states is displayed by using a colorbar on them. This quantity indicates pseudo-spin polarization. It is seen that at the point $\beta_2$ which corresponds to QAHI phase, the edge states are not perfectly pseudo-spin polarized. The reason is originated to the presence of SIA potential in this point which causes to have a mixture of pseudo-spin states in the edge modes. So the pseudo-spin polarization of the edge states will be deviated from $\pm1$ even if we still stay at QAHI phase. Moreover, the point marked by $\beta_4$, is in QPHI phase which has two helical pseudo-spin edge modes, so the expectation value of the pseudo-spin operator is nearly zero. In this phase, we have no fully helical edge modes, since SIA potential is non-zero. For the point $\beta_2$, The eigen value of $\tilde{\tau}_z$, is in the range of $0<\langle\tilde{\tau}_z\rangle<1$, while at the point $\beta_1$, we have fully pseudo-spin polarization $\langle\tilde{\tau}_z\rangle=1$. 

The influence of increasing the radiation amplitude is traced back into the rescaling of the Fermi velocity of the edge states as seen in Eq.~\ref{Photo_Hamiltonian2}. The coefficient of Fermi velocity in the mentioned equation, i.e. $ (1-2A'\Delta_1 \tilde{\tau}_z)$ determines the velocity of the edge states. In QPHI phase region in which SIA potential is zero, the Fermi velocity for each pseudo-spin is different. This anisotropy in Fermi velocity of the helical edge modes, leads to asymmetric Josephson effects when a two dimensional TI with inversion symmetry breaking is mediated between two superconductors\cite{asymmetric_velocity}.

\subsubsection{Pseudo-spin Chern Number} \label{S3_ba}
As we mentioned before, pseudo-spin Chern number can be defined even if pseudo-spin is not a conserved quantity \cite{Prodan,Sheng,spinchernnumber,spinchernnumberTI}. This Chern number is protected by two gaps: the band gap and the pseudo-spin gap. The latter is the gap in the spectrum of $\mathcal{S}=\mathcal{P}\tilde{\tau}_z\mathcal{P}$ where $\mathcal{P}$ is a projector onto valence bands. Suppose that we have two valence bands which are called as $|v1\rangle$ and $|v2\rangle$.

\begin{equation}
\mathcal{S}=\left(\begin{array}{cc}{\langle}v1|\tilde{\tau}_z|v1\rangle & {\langle}v1|\tilde{\tau}_z|v2\rangle\\ {\langle}v2|\tilde{\tau}_z|v1\rangle&{\langle}v2|\tilde{\tau}_z|v2\rangle \\\end{array} \right)
\label{pspec}
\end{equation}

\begin{figure}
\includegraphics[width=0.9 \linewidth]{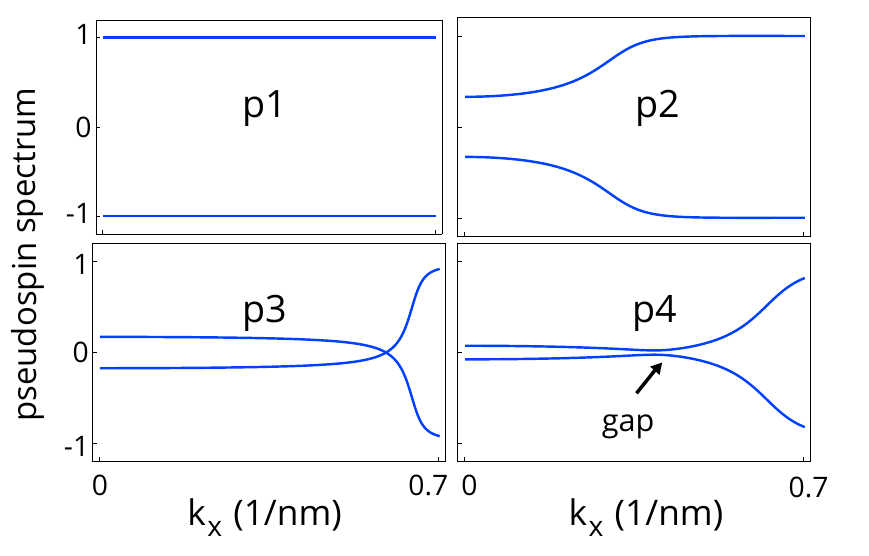} 
\caption{The eigenvalues of $\mathcal{S}$ operator defined in Eq.~\ref{pspec} for points $p1-p4$ shown in Fig.~\ref{avdiagram}(f) .}
\label{ps}
\end{figure}

So in order to define pseudo-spin Chern number, there should always be a gap in the pseudo-spin spectrum. We checked the satisfaction of this condition in distinct phases. When $V_{sia}=0$, it is obvious from Eq. \ref{eq:hamiltotalchiral} that pseudo-spin spectrum is fully gapped and the eigenvalues of $\mathcal{S}$ operator are equal to $\pm1$. In Fig.\ref{avdiagram}(c-f), when $\mathcal{A}=0$ in the white (NI) region, the pseudo-spin spectrum is gappless. As long as we checked numerically, in other regions, when the band gap is finite, the pseudo-spin spectrum is also gapped. This behaviour can be seen in Fig.\ref{ps} where the $\mathcal{S}$ spectrum for the special points $p1-p4$ shown in Fig. \ref{avdiagram}(f), is presented as a function of $k_x$, ($k_y=0$ in this plot). Turning on a small $\mathcal{A}$ can open a gap in $\mathcal{S}$ spectrum when it is gapless (see Figs.\ref{ps}(p3,p4)). 

While the $\mathcal{S}$ spectrum is gapped, the pseudo-spin Chern number can be calculated for each eigenfunction of $\mathcal{S}$ by numerical methods. The result is shown on each phase in Fig. \ref{avdiagram}. 

\subsection{Linear Polarization}\label{S4}
For the sake of completeness, we try to determine topological phases of a thin topological insulator irradiated by a high-frequency light with {\it linear} polarization. As we mentioned before, we should use the BW or vV expansion up to the second order perturbation in terms of inverse frequency powers. In this section, it is assumed that the z-axis magnetization is zero ($M_z=0$). As proved in the appendix, the first order correction is zero and therefore, up to the second order of $1/\Omega$, the effective Hamiltonian would be derived as

\begin{equation}
\begin{aligned}
&H^{\text{linear}} (\textbf{k})=\left(\begin{array}{cc} \tilde{h} & q \\ q & \tilde{h}^T \\\end{array} \right) \\
&\tilde{h}=\left(\begin{array}{cc} \tilde{\Delta}_0+\Delta_x k_x^2+\Delta_y k_y^2 & i A_x k_x +A_y k_y+A_3 k_-^2 k_y \\ -i A_x k_x +A_y k_y+A_3 k_+^2 k_y & -(\tilde{\Delta}_0+\Delta_x k_x^2+\Delta_y k_y^2) \\\end{array} \right) \\
&q=\left(\begin{array}{cc}V'' k_y & \tilde{V}+V' k_y^2 \\ \tilde{V}+V' k_y^2 & -V'' k_y \\\end{array} \right)
\label{linhamil} 
\end{aligned}
\end{equation}

\begin{figure}
\includegraphics[width= \linewidth]{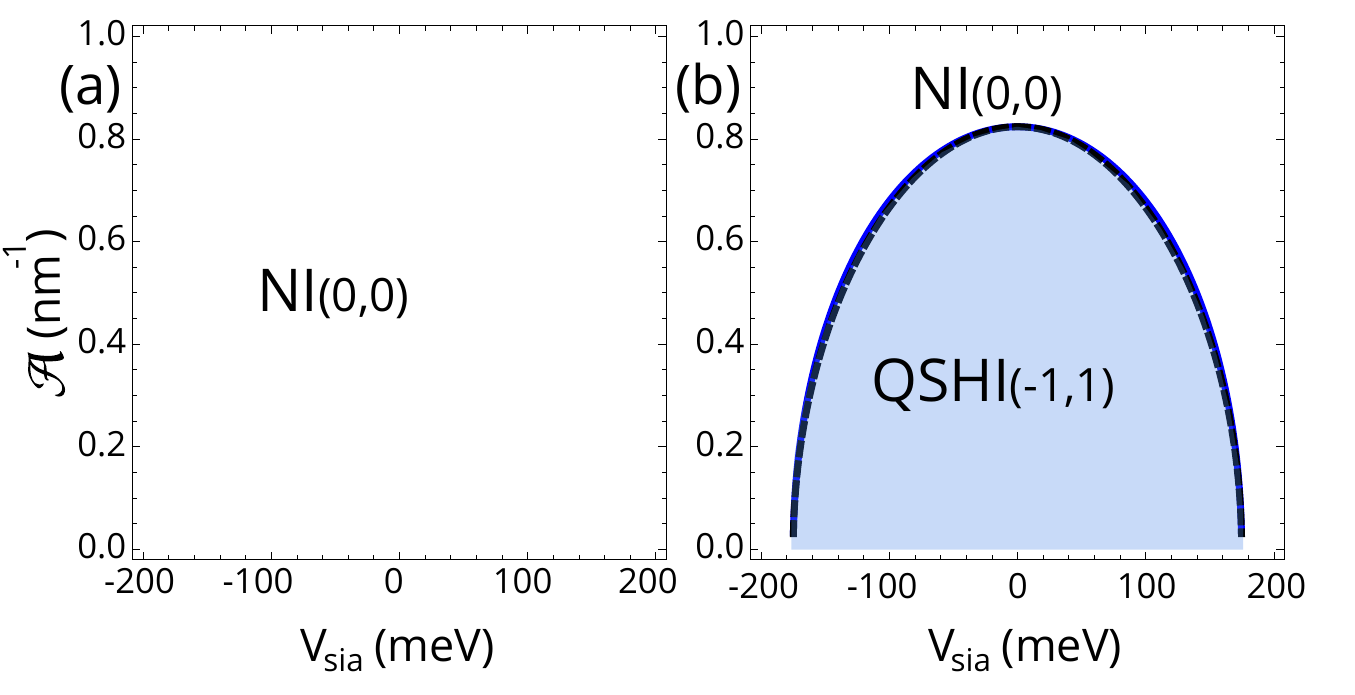} 
\caption{The phase diagram for linear polarization in $\mathcal{A}-V_{sia}$ plane when $M_{z}=0$. The panels a (b) are for $\Delta_0\Delta_1>0 (\Delta_0\Delta_1<0)$. The blue (black dashed) line corresponds to gap-closing at $k_x=0 (k_y=0)$ respectively.}
\label{linfig}
\end{figure}

Where $k_{\pm}=k_x \pm i k_y$ and other new parameters are defined in Appendix. \ref{A1}. It should be mentioned that illumination of the linearly polarized light does not break time reversal symmetry, however the circularly polarized light and also magnetization break it. So in the presence of time reversal symmetry, the Chern number would be zero and one can compute $Z_2$ invariant to find out whether the associated phase is QSHI or NI. To obtain the phase diagram of this Hamiltonian, at first, we search for the gap closings. In the absence of $V_{\text{sia}}$:\\
The gap closes at the $\Gamma$ point if $\tilde{\Delta}_0=0$. The gap would also close at $\tilde{\Delta}_0+\Delta_x k_x^2+ \Delta_y k_y^2=0$ provided that $k_x(A_x-2A_3 k_y^2)=0 $ and $k_y (A_y+A_3 k_x^2-A_3 k_y^2 )=0$. However, it should be noted that these three conditions for the gap closing will not be satisfied simultaneously in our desired range of parameters.\\
In the presence of $V_{sia}$:\\
The gap closes at $k_x=k_y=0 $ if $\tilde{V}^2+\tilde{\Delta}_0^2=0$. Nevertheless, $\tilde{V}$ and $\tilde{\Delta}_0$ should vanish simultaneously which can not happen for $V_{sia}\neq0$. The condition for out of $\Gamma$ gap-closing is difficult to derive analytically, so let us consider some special cases. For instance, the gap is closed at $k_y=0$, $k_x^2=-\tilde{\Delta}_0/\Delta_x >0$ if $A_x \sqrt{\frac{-\tilde{\Delta}_0}{\Delta_x}}=\pm \tilde{V}$. This kind of gap closing is shown by the black dashed line displaying in the phase diagrams shown in Fig.~\ref{linfig}. Phase diagram for the case $\Delta_0\Delta_1>0 (\Delta_0\Delta_1<0)$ corresponds to Fig.~\ref{linfig}a(b). One can search for the other out of $\Gamma$ gap-closing points however we checked that in the favorite range of parameters, they represent nearly the same boundaries for the phase transitions. For example the gap can be closed at $k_x=0$ and $k_y^2\Delta_y-k_y V''+\tilde{\Delta}_0=0$ when $-A_yk_y+k_y^2 (A_3 k_y+V')+\tilde{V}=0$, which is shown by the blue line represented in Fig.~\ref{linfig}. From the phase diagrams of the case (i) and (ii) which are shown in Fig.~\ref{linfig}(a),(b), it is seen that by variation in the field amplitude, a phase transition occurs from quantum spin Hall insulator (QSHI) to NI for the case (ii).

It is notable that we have used a series expansion of high frequencies to extract the effective Hamiltonian and truncated it to the second order. This truncation is not completely reliable for large amplitudes of the drive $\mathcal{A}>1.5$. To re-derive our results, we have also diagonalized the full Floquet Hamiltonian with the dimension up to the convergence of the result\cite{rudnerrev} and found that there is a good agreement between the expansion and Floquet Hamiltonian result in range of parameters presented in Fig.~\ref{linfig}. Actually there are also additional phase transition lines near $\mathcal{A}=3.89 (4.06)$ for the case i(ii), which are not displayed in the phase diagram mentioned in Figs. \ref{linfig}(a,b). There is a comparison between the Hamiltonian extracted by the series expansion and the full Floquet Hamiltonian which is presented in Appendix. \ref{A2} 

\section{Experimental Realization} \label{S3_c}
Experimental realization of the phases is important and challeging. The main challenge to this end may be the heating effect which is caused by intense irradiation on the system. Thanks to high frequency regime which opens up a solution for avoiding the heating. Indeed, the heating rate is exponentially suppressed in $\omega$ at a driving frequency larger than any energy change arising from single scattering events in the system \cite{rudnerrev}. 

If the time scale over which the heating process is running is extremely long, on short or intermediate time scales, the driven system may be stable in a nearly time-periodic quasi-steady state. Any Floquet state including topological phases emerges during this time scale. Therefore, in experiment, to implement Floquet engineering in solids induced by high driving amplitude, we are forced to use ultra-short laser pulses which its pulse duration is shorter than the time scale for the heating. The photon energy which is used in this work, corresponds to a period of 4 fs which is enough to have a quasi-steady time-periodic state. As mentioned in introduction, recently, light-induced quantum anomalous Hall effect in graphene has been reported by means of an ultra-fast transport setup \cite{mciver} where a 500 fs laser pulse is used to induce this effect.

\begin{figure}
\includegraphics[width=0.8 \linewidth]{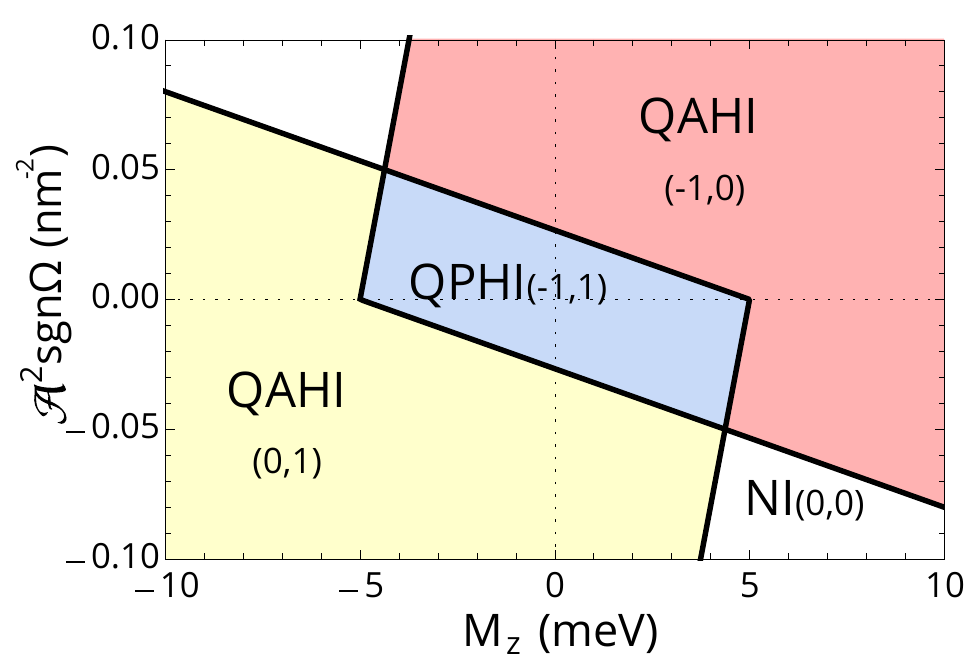} 
\caption{The phase diagram in $\mathcal{A}^2 sgn(\Omega)-M_z$ plane when $V_{sia}=0$. The thin film parameters are considered to be $\Delta_0=5$~meV and $\Delta_1=-10$~eV\AA$^2$ and the phton energy is $\hbar \Omega=1$~eV.}
\label{d5}
\end{figure}

On the other hand, the required amplitude of the drive for observing a phase transition, can be reduced by for example changing a film thickness which leads to a smaller band gap. The phase diagram for a film with smaller gap, $\Delta_0=5$~meV is presented in Fig.\ref{d5}. In this diagram, it is seen that the phase transitions occur at lower amplitudes of the drive. One can also tune $M_z$ and/or $V_{sia}$ such that the system becomes nearly gaplees and so by means of a very small amplitude of the drive, it is feasible to occur these phase transitions. If the scaled strength is equal to $\mathcal{A}=0.2/nm$ then the amplitude of the electric field would be $E_0=\frac{\mathcal{A} \hbar\Omega}{e}=\frac{0.2/nm*1 eV}{e}=2*10^8\frac{V}{m}$ which is a reasonable amplitude.

To avoid partially the heating problem, one way is to construct the system Hamiltonian artificially in ultra-cold atom gases in which lattice parameter could be enough large for lowering the band width and as a consequence, lower frequecy is required for guaranteeing the off-resonant regime, for example in graphene.  

\section{Conclusion}\label{S5}
In this work, we investigate the influence of high frequency circularly and linearly-polarized light on topological insulator thin films, while simultaneously the effect of structure inversion asymmetry,magnetic dopants and thickness engineering are also considered. The interplay between these perturbations gives rise to a rich phase diagram, where NI, QAHI, QPHI and QSHI phases are notable. We showed that in the range of considered parameters, turning on SIA potential can change the quantum phase of the system to QPHI and/or NI if it exceeds a critical potential. Moreover, SIA potential breaks pseudo-spin polarization and causes the edge states to be not fully pseudo-spin polarized. We derive a different effective Floquet Hamiltonian from what is reported in Ref.~\cite{shelykh} and show that $V_{sia}$ can change the phases represented in the mentioned reference provided that it exceeds a critical potential. For the sake of completeness, we also derive an effective Floquet Hamiltonian for our studied system (a thin film of topological insulator) driven by a linearly polarized light. The phase diagram shows that a QSHI phase can be inverted to NI phase only by altering $V_{sia}$. 
\appendix
\section{Effective Hamiltonian irradiated by linear-polarized light}\label{A1}
The effect of linearly polarized light on matter can be considered by the Peierls substitution $\textbf{k} \rightarrow \textbf{k}+e \textbf{A}/\hbar$ where the vector potential is equal to $\textbf{A}=A_0 (0,cos\Omega t)$. Then the Fourier components of the time-dependent Hamiltonian are simply calculated as (we set $\hbar=1$ in the appendix) 
\begin{equation}
\begin{aligned}
H^0=&H^{\text{dark}} +\Delta_1\mathcal{A}^2/2 \tilde{\sigma}_z \\
H^1=&H^{-1}=v_f \mathcal{A} \tilde{\sigma}_x/2+\mathcal{A} k_y \Delta_1 \tilde{\sigma}_z \\
H^2=&H^{-2}=\mathcal{A}^2\Delta_1 \tilde{\sigma}_z/4
\label{h0} 
\end{aligned}
\end{equation}
The first correction of the Hamiltonian in inverse powers of frequency ($1/\Omega$) is proportional to the commutation relation $[H^{-1},H^1]=0$, which is zero. The second order correction can be calculated by means of the BW or vV expansion \cite{bw}. After a straightforward calculation, we find that
\begin{equation}
\begin{aligned}
H^{(2)}_{BW}=& \big([H^1,[H^1,H^2]] +2 (H^1H^0H^1-H^1H^1H^0) \\
& + (H^2H^0H^2-H^2H^2H^0)/2 \big)/\Omega^2 \\
H^{(2)}_{vV}=&\big( [H^1,[H^1,H^2]] + [H^1,[H^0,H^1]]\\
&+[H^2,[H^0,H^2]]/4 \big)/\Omega^2
\label{2hs} 
\end{aligned}
\end{equation}
These two expressions yield the same result for our model, however they are different from what is reported in Eq. (A6) of Ref. \cite{shelykh}. By substituting Eq. \ref{h0} in Eq. \ref{2hs} and adding the term $H_0$ we obtain the effective Hamiltonian of Eq.~\ref{linhamil} with the following parameters
{\setstretch{0.4}
\begin{equation}
\tilde{\Delta}_0\equiv\Delta_0+(\frac{\Delta_1}{2}-\frac{v_f^2 \Delta_0}{\Omega^2})\mathcal{A}^2-(\frac{\Delta_1}{4}\frac{v_f^2}{\Omega^2})\mathcal{A}^4
\end{equation}
\begin{equation}
\frac{\Delta_x}{\Delta_1}\equiv 1-\frac{v_f^2}{\Omega^2}\mathcal{A}^2 
\end{equation}
\begin{equation}
\frac{\Delta_y}{\Delta_1} \equiv 1+\frac{v_f^2}{\Omega^2}\mathcal{A}^2
\end{equation}
\begin{equation}
\frac{A_x}{v_f}\equiv 1-\frac{v_f^2}{\Omega^2}\mathcal{A}^2-\frac{1}{16}\frac{\Delta_1^2}{\Omega^2}\mathcal{A}^4
\end{equation}
\begin{equation}
\frac{A_y}{v_f}\equiv 1+2\frac{\Delta_0\Delta_1}{\Omega^2}\mathcal{A}^2+\frac{7}{16}\frac{\Delta_1^2}{\Omega^2}\mathcal{A}^4
\end{equation}
\begin{equation}
\frac{A_3}{v_f}\equiv 2\frac{\Delta_1^2}{\Omega^2}\mathcal{A}^2 
\end{equation}
\begin{equation}
V''\equiv 2\frac{v_f\Delta_1 V_{sia}\mathcal{A}^2}{\Omega^2} 
\end{equation}
\begin{equation}
\tilde{V}\equiv V_{sia}-\frac{1}{16}\frac{V_{sia} \Delta_1^2}{\Omega^2}\mathcal{A}^4 
\end{equation}
\begin{equation}
V'\equiv-4\frac{V_{sia} \Delta_1^2 \mathcal{A}^2}{\Omega^2}
\end{equation}}

\begin{figure}
\includegraphics[width=0.8 \linewidth]{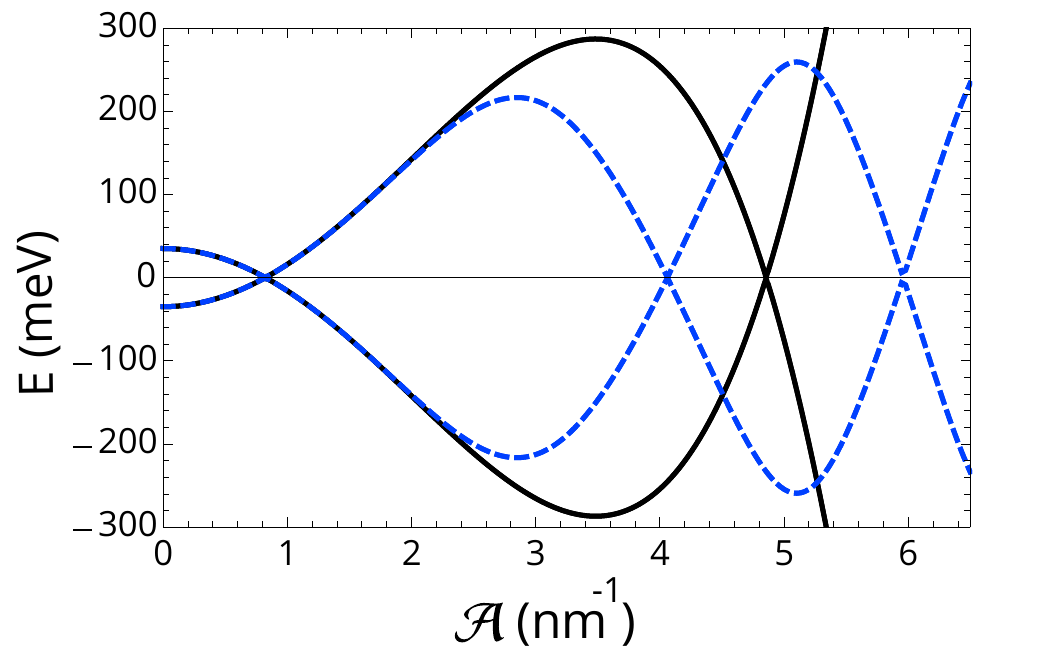} 
\caption{Energy bands at $\Gamma$ point as a function of the amplitude of the drive for $V_{sia}=0$ and $\Delta_0\Delta_1<0$. The black (dashed blue) line is derived by the expanded Hamiltonian (Floquet Hamiltonian truncated up to the convergence of the results), respectively.}
\label{range}
\end{figure}


\section{comparison between expanded Hamiltonian and non-perturbative method}\label{A2}
The truncation of the expanded Hamiltonian to the second order (Eq.~\ref{2hs} ) is not sufficient for large amplitudes of the drive. This can be measured by diagonalizing the Floquet Hamiltonian (truncated up to the dimensions which guarantee the convergence of the result) \cite{rudnerrev}. The valence and conduction energy bands at $\Gamma$ point are ploted in terms of light amplitude in Fig.~\ref{range} for $V_{sia}=0$ and $\Delta_0\Delta_1<0$. In this figure, the energy bands calculated by the expanded Hamiltonian (black lines) are compared with those bands which are calculated by the full Floquet Hamiltonian (dashed lines). It is apparent that for amplitudes larger than $\mathcal{A}=2$, the expanded Hamiltonian results deviate from the exact results and so the expanded Hamiltonian is not applicable for large value of the drive. The expanded Hamiltonian method yields reasonable results in the range of parameters considered in Figs.~\ref{linfig}.

\end{document}